\documentclass[aps,pra,reprint,superscriptaddress,noeprint]{revtex4-2}

\usepackage{amsmath}
\usepackage{graphicx}
\usepackage{hyperref}

\begin{document}

\title{Optimizing spontaneous parametric down-conversion sources for boson sampling}
\author{R.~van~der~Meer}
\email[]{r.vandermeer-1@utwente.nl}
\author{J.J.~Renema}
\affiliation{COPS, MESA+ Institute for Nanotechnology, University of Twente, PO Box 217, 7500 AE Enschede, The Netherlands}
\author{B.~Brecht}
\author{C.~Silberhorn}
\affiliation{Integrated Quantum Optics, Paderborn University, Warburger strasse 100, 33098 Paderborn, Germany}
\author{P.W.H.~Pinkse}
\affiliation{COPS, MESA+ Institute for Nanotechnology, University of Twente, PO Box 217, 7500 AE Enschede, The Netherlands}
\date{\today}

\begin{abstract}
\textbf{\noindent An important step for photonic quantum technologies is the demonstration of a quantum advantage through boson sampling. In order to prevent classical simulability of boson sampling, the photons need to be almost perfectly identical and almost without losses. These two requirements are connected through spectral filtering, improving one leads to a decrease of the other. A proven method of generating single photons is spontaneous parametric downconversion (SPDC). We show that an optimal trade-off between indistinguishability and losses can always be found for SPDC. We conclude that a 50-photon scattershot boson-sampling experiment using SPDC sources is possible from a computational complexity point of view. To this end, we numerically optimize SPDC sources under the regime of weak pumping and with a single spatial mode.}
\end{abstract}

\maketitle
\begin{figure}
\includegraphics{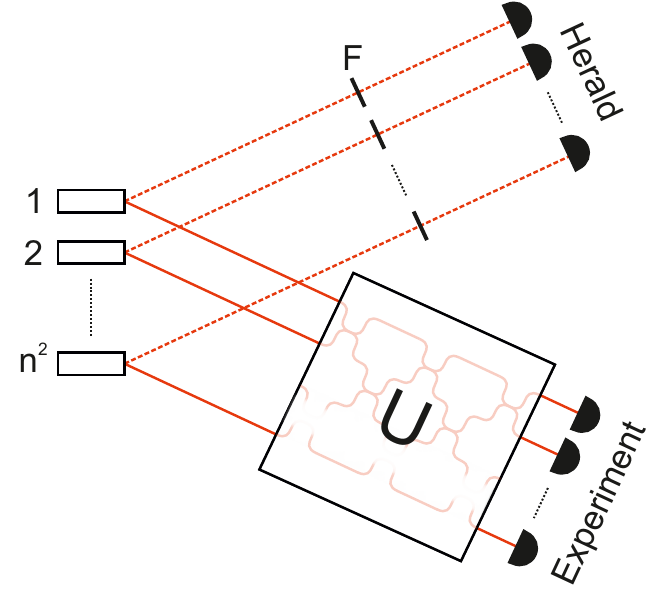}%
\caption{\label{fig:fig1BosonSampling}A $n$-photon scattershot boson-sampling experiment has $n^2$ heralded single-photon sources. Each source can send a photon to one of the input modes of the interferometer $U$. The other photon (dashed) is filtered (F) and is used as a herald.}
\end{figure}

The next milestone in photonic quantum information processing is demonstrating a quantum advantage \cite{harrow_2017_Nature,arute_2019_Nature}, i.e. an experiment in which a quantum optical system outperforms a classical supercomputer. This can be achieved with boson sampling \cite{Aaronson_2013}. The aim of boson sampling is, for a given input configuration of photons, to provide a sample of the output configuration from a arbitraryunitary transformation. A photonic quantum device which implements this consists of multiple photon sources, a large passive interferometer and single-photon detectors as shown in Fig. \ref{fig:fig1BosonSampling}. This is believed to be easier to implement than a universal quantum computer and resulted in a surge of experiments \cite{broome_2013_Science,spring_2013_Science,tillmann_2013_NatPhotonics,crespi_2013_NatPhotonics,wang_2017_Nat.Photonics,zhong_2018_Phys.Rev.Lett.,paesani_2018_ArXiv181203158Phys.Physicsquant-Ph,wang_2019_Phys.Rev.Lett.}. These experiments require many almost identical photons and practically no losses.


Spontaneous parametric downconversion (SPDC) sources are a well-known method of generating single photons. A major drawback of building an $n$-photon SPDC source is the probabilistic generation of the photon pairs, meaning that generating $n$ photons simultaneously will take exponentially long. Scattershot boson sampling improves on this by enabling the generation of $n$ photons in polynomial time using $\sim n^2$ sources in parallel \cite{latmiral_2016_NewJPhys}. The photons, however, still need to be sufficiently identical.


A way to improve the photon indistinguishability is spectral filtering. Unfortunately, this comes at the cost of losses. Losses, too, are detrimental to multiphoton interference experiments as they exponentially increase the experimental runtime \cite{rohde_2012_PhysRevA}. Finding an optimal trade-off between losses and distinguishability is a nontrivial task.

Previous work on optimizing the spectral filtering SPDC sources focused on a trade-off between spectral purity and symmetric heralding efficiency \cite{meyer-scott_2017_Phys.Rev.A}. Other work on designing SPDC sources has studied optimal focusing parameters for bulk crystal sources and pump beam parameters \cite{bennink_2010_Phys.Rev.A}, and phase-matching functions \cite{graffitti_2018_Phys.Rev.A}. However, the design of SPDC sources specifically for boson sampling remains an open question as the optimal trade-off between losses and indistinguishability has not been studied so far.

Recently a new classical approximation algorithm for noisy boson sampling was suggested which incorporates both losses and distinguishability \cite{renema_2018_ArXiv}. This algorithm gives a lower bound to the amount of imperfections that can be tolerated in order to still achieve a quantum advantage. More importantly, since it incorporates both imperfections, it can be used to trade-off distinguishability and losses.

In this work, we investigate the design of SPDC sources for scattershot boson sampling from a complexity theory point of view. The model of \cite{renema_2018_ArXiv} is used to find the optimal source and filter parameters for a boson-sampling experiment. From this we determine a minimal overall transmission efficiency which places a lower bound on the transmission by other experimental components. We target, by convention, a 50-photon boson-sampling experiment \cite{neville_2017_Nat.Phys.}. 

Three SPDC crystals are considered: potassium titanyl phosphate (ppKTP), $\rm{\beta}$-barium borate (BBO) and potassium dihydrogen phosphate (KDP). KTP is a popular choice since it has symmetric group velocity matching at telecom wavelengths \cite{kim_2006_Phys.Rev.A,weston_2016_Opt.ExpressOE}, which is favorable for obtaining pure states. The photon generation rates of KTP sources are high as it  uses periodic poling. Moreover, periodic poling allows Gaussian-shaped phase-matching functions by means of Gaussian apodization \cite{branczyk_2011_Opt.ExpressOE,dixon_2013_Opt.ExpressOE,chen_2017_Opt.ExpressOE,graffitti_2018_Optica}. The second crystal, BBO, is known for generating the current record number of photons \cite{kwiat_1995_PhysRevLett,zhong_2018_Phys.Rev.Lett.} and also generates photons at telecom wavelength. However, it has asymmetric group velocity matching, resulting in a reduced spectral purity. Finally, the last crystal we consider is KDP. KDP sources, which generate photons at $830\,$nm, are known to generate one of the highest purity photons without filtering \cite{mosley_2008_Phys.Rev.Lett.}.

Our calculations consider Gaussian-shaped pulses to pump the SPDC process in a collinear configuration. We assume the existence of only one spatial mode and do not take into account focusing effects. This is a valid assumption for both waveguide sources as well as for bulk sources without focusing. Focusing increases the number of spatial modes and hence affects the spectral purity \cite{bennink_2010_Phys.Rev.A}. Furthermore, higher-photon-number states are ignored, which is reasonable given the existence of photon-number-resolving detectors \cite{rosenberg_2005_Phys.Rev.A}.

\section{\label{Theory}Theory}
\subsection{SPDC sources}
\begin{figure}
\includegraphics[width=0.95\columnwidth]{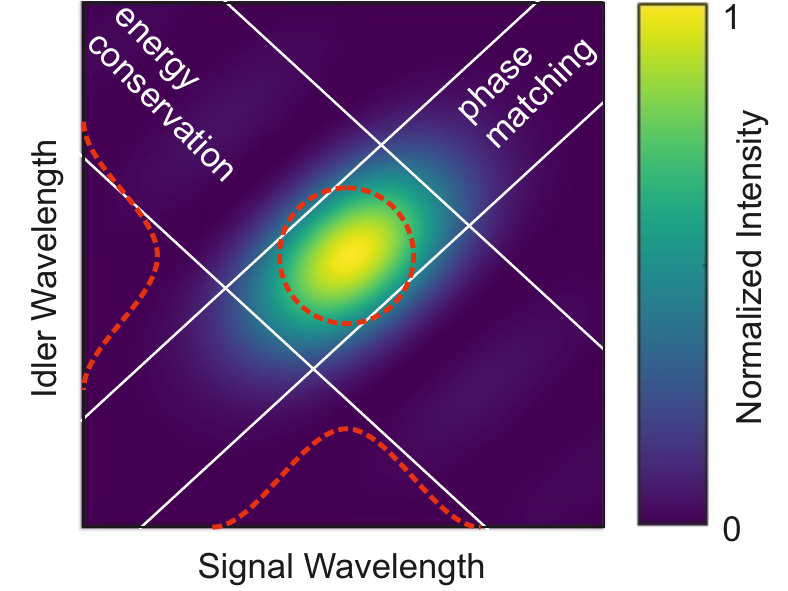}%
\caption{\label{fig:fig2_JSA}An example of a joint spectral intensity (JSI). The red dashed line shows the Gaussian filter for both the signal and idler photon. The (anti)diagonal white lines denote the region which satisfies phase matching (energy conservation).}
\end{figure}
SPDC sources turn a pump photon into two down-converted photons, and hence produce photons in pairs. For Type-II SPDC, the two photons from the pair each emerge in a separate mode. Traditionally these modes are referred to as signal and idler. The SPDC process can be understood by considering energy conservation $\hbar\omega_{\rm p} = \hbar\omega_{\rm s} + \hbar\omega_{\rm i}$ as well as momentum conservation $\vec{k_{\rm p}}= \vec{k_{\rm s}} + \vec{k_{\rm i}}$, where p, s and i denote the pump, signal and idler photons, respectively. Momentum conservation can be tweaked by quasi phase matching by either periodic or apodized poling. Both energy and momentum conservation only allow certain wavelength combinations and together they specify the spectral-temporal properties of the two-photon state \cite{grice_1997_PhysRevA}.

Birefringence results in an asymmetry between the signal and idler photon. This leads to spectral-temporal correlations between the two. Such correlations reduce the spectral purity $P_{\rm x} = \rm{Tr}( \rho_{\rm x}^2)$, where $\rho_{\rm x}$ is the reduced density matrix of photon x. When no correlations exist, the photon state is factorizable and the photons are spectrally pure \cite{URen_laserPhysics_2005}.

A visual representation of the two-photon state is shown in Fig. \ref{fig:fig2_JSA}. The spot in the center indicates that the two-photon state with what probability the photons are in this region of the frequency space. This probability is also referred to as the joint spectral intensity (JSI), which is related to the joint spectral amplitude (JSA) by $\rm{JSI} = |\rm{JSA}|^2$. The JSA describes the wavefunction of the photon pair as a function of the wavelength of the photons and follows from energy and momentum conservation. The factorizability of the JSA determines the spectral purity of the source. 

We now proceed with a mathematical description of the JSA, which follows from energy and momentum conservation. The energy conservation $\alpha(\omega_{\rm s},\omega_{\rm i})$ function is a Gaussian pulse with a center wavelength $\omega_{\rm p}$ and bandwidth $\sigma_{\rm p}$:
\begin{equation}
\alpha(\omega_{\rm s},\omega_{\rm i}) = \exp\biggl(\frac{-(\omega_{\rm s}+\omega_{\rm i}-\omega_{\rm p})^2}{4\sigma_{\rm p}^2}\biggr).
\end{equation}
The phase-matching function for a periodically poled crystal is given by:
\begin{equation}
\phi(\omega_{\rm s},\omega_{\rm i}) = {\rm sinc}\biggl( \frac{k_{\rm p}-k_{\rm s}-k_{\rm i}-\frac{2\pi}{\Lambda}}{2}L\biggr),
\label{eq:eqPMsince}
\end{equation}
with $L$ the length of the nonlinear crystal and $\Lambda$ the poling period. Another type of quasi phase matching exists, which is Gaussian apodization \cite{branczyk_2011_Opt.ExpressOE}
\begin{equation}
\phi_{\rm G}(\omega_{\rm s},\omega_{\rm i}) = \exp{\biggl(-\frac{\gamma \Delta k^2 L^2}{4}\biggr)},
\end{equation}
where $\gamma \approx 0.193$, such that the width of this phase-matching function equals that of Eq. \ref{eq:eqPMsince}. The parameter $\Delta k$ denotes the phase mismatch and $L$ again the crystal length. The energy conservation function together with the appropriate phase-matching function give the JSA:
\begin{equation}
f(\omega_{\rm s},\omega_{\rm i}) = \alpha(\omega_{\rm s},\omega_{\rm i}) \phi(\omega_{\rm s},\omega_{\rm i}).
\end{equation}
The two-photon state corresponding to this JSA can give rise to distinguishability. This can be mitigated by spectral filtering. The overall two-photon state after filtering can now be written as:
\begin{equation}
|\psi \rangle = \int\int d\omega_{\rm s} d\omega_{\rm i} f(\omega_{\rm s},\omega_{\rm i}) F_{\rm s,i}(\omega_{\rm s},\omega_{\rm i}) |1_{\rm s}\rangle |1_{\rm i}\rangle,
\end{equation}
where $F_{\rm s,i}(\omega_{\rm s},\omega_{\rm i})$ denotes a possible filter function on the signal and/or idler photon. For simplicity, we ignore the vacuum and multiphoton states. 

The spectral purity of the photon pair can be found with a Schmidt decomposition of the JSA \cite{law_2000_Phys.Rev.Lett.,eberly_2006_LaserPhys.}. From this follows a Schmidt number $K$ which determines the spectral purity
\begin{equation}
P = \frac{1}{K}.
\end{equation}
Physically, $K$ is the effective number of modes that is required to describe the JSA (e.g., see \cite{christ_2011_NewJ.Phys.}). When $K=1$ the photon pair is factorizable. In this case, detecting a photon as herald leaves the other photon in a pure state. In Fig. \ref{fig:fig2_JSA} this would manifest itself such that the JSA becomes aligned with the axes. In case $K>1$, detecting one photon leaves the other photon in a mixed state of several modes. Hence, the remaining photon has a lower spectral purity.

It is possible to improve the spectral purity by filtering the photons. The effect of filtering can be understood as overlaying the filter function over the JSA. This is shown with the dashed lines in Fig. \ref{fig:fig2_JSA}. A well-chosen filter removes the frequency correlations between the photons, but inevitably introduces losses, which in turn are detrimental for boson-sampling experiments.

\subsection{Classical simulation of boson sampling with imperfections}
The presence of imperfections such as losses \cite{garcia-patron_2019_Quantum} and distinguishability \cite{renema_2018_Phys.Rev.Lett.} of photons reduces the computational complexity of boson sampling. Classical simulation algorithms of boson sampling upper bound the allowed imperfections. These classical simulations approximate the boson sampler outcome with a given error. 

We now present the model of \cite{renema_2018_ArXiv}. This model approximates an imperfect $n$-photon boson sampler where $n-m$ photons are lost, by describing the output as up to $k$-photon quantum interference ($0\leq k\leq m$) and at least $m-k$ classical boson interference. Furthermore, this formalism naturally combines losses and distinguishability into a single simulation strategy, thereby introducing an explicit trade-off between the two. In this model, the error bound $E$ of the classical approximation is given by
\begin{equation}
E < \sqrt{ \frac{\alpha^{k+1}}{1-\alpha}}.
\label{eq:eqBound}
\end{equation}
The parameter $\alpha$ which we will refer to as the 'source quality' is given by
\begin{equation}
\alpha = \eta x^2,
\label{eq:eqAlpha}
\end{equation}
with $\eta=m/n$ denotes the transmission efficiency per photon. Losses in different components are equivalent, so different losses can be combined into a single parameter $\eta$ \cite{oszmaniec_2018_NewJ.Phys.}. The average overlap of the internal part of the wave function between two photons is given by $x=\langle \psi_{\rm i} | \psi_{\rm j} \rangle$ (i$\neq$j). Therefore $x^2$ is the visibility of a signal-signal Hong-Ou-Mandel interference dip \cite{hong_1987_PhysRevLett}. This indistinguishability equals the spectral purity.

This model allows for optimizing the SPDC configuration by optimizing the source quality of Eq. \ref{eq:eqAlpha}, which effectively trades-off the losses and distinguishability. Furthermore, from Eq. \ref{eq:eqBound} the maximal number of photons $k$ can be calculated by specifying a desired error bound.

\section{\label{Methods}Methods}
In order to find the best SPDC configuration for a selection of crystals, we run an optimization over the SPDC settings to maximize $\alpha$ while varying the filter bandwidth. Since we consider collinear SPDC, the optimization parameters are the crystal length $L$ and the pump bandwidth $\sigma_{\rm p}$. Note that these parameters determine the shape of the JSA and therefore the separability. The pump center wavelength is set such that group velocity dispersion is matched \cite{keller_1997_Phys.Rev.A,grice_2001_Phys.Rev.A,gerrits_2011_Opt.ExpressOE,mosley_2008_Phys.Rev.Lett.}. From our numerical calculations we observe that the optimization problem appears to be convex over the region of the parameter space of interest. We note that the optimization parameters are bounded, e.g., the crystal length cannot be negative. A local optimization routine (L-BFGS-B, Python) was used.

\begin{figure*}
\includegraphics{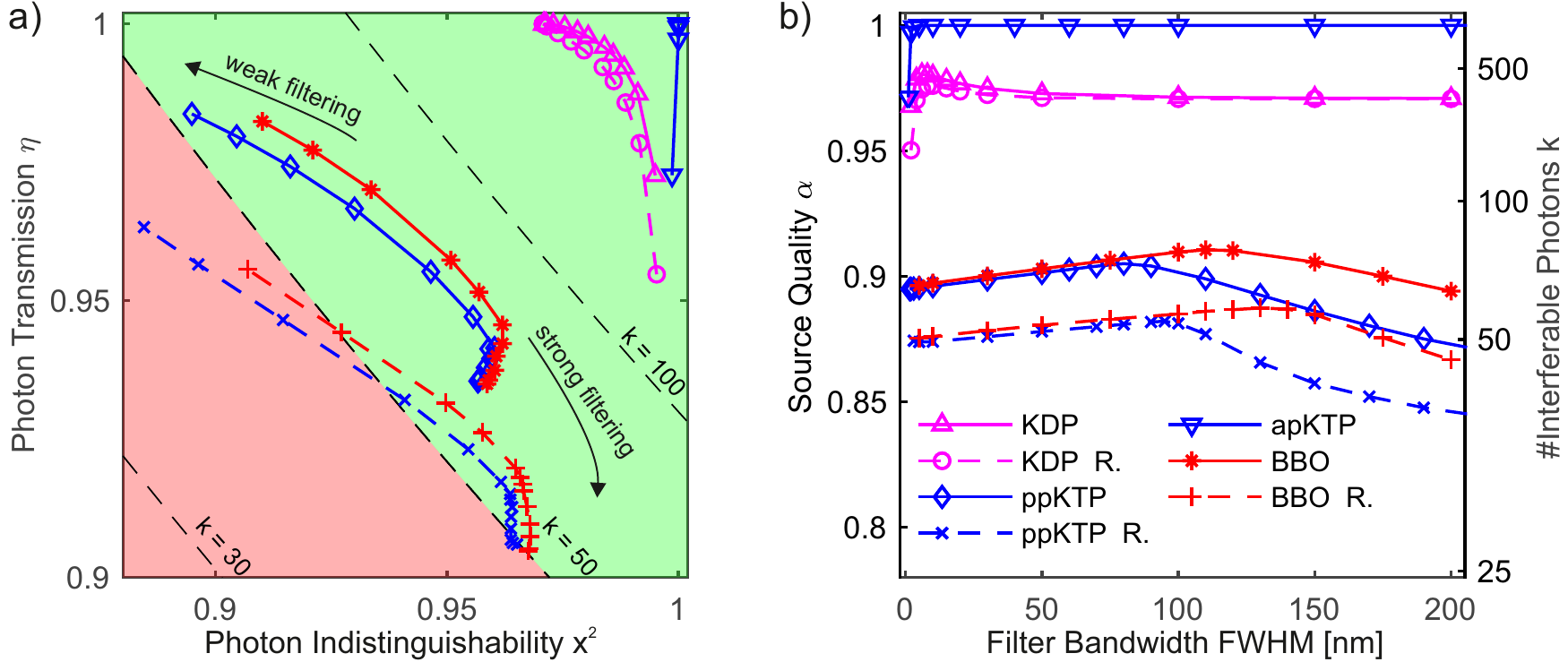}%
\caption{\label{fig:fig3_results}a) The transmission efficiency per photon $\eta$ and indistinguishability $x^2$ corresponding to the ideal SPDC settings at different filter bandwidths for different crystals (see legend in b). The dashed lines are isolines, indicating how many photons $k$ can be used for a boson-sampling experiment. The indistinguishability and transmission efficiency together result in the source quality factor $\alpha = x^2 \eta$. b) The values of $\alpha$ and the corresponding number of photons $k$ (right axis) as function of the filter bandwidth. In the legend R. denotes a rectangular filter, otherwise a Gaussian filter was used.}
\end{figure*}

The source quality $\alpha$ can be calculated from the JSA. The JSA was calculated numerically by discretizing the wavelength range of interest \cite{suppMaterial}. The wavelength range was chosen to include possible side lobes of the sinc phase-matching function. The spectral purity is calculated from the discretized JSA using a singular value decomposition (SVD) \cite{mosley_2008_NewJ.Phys.}. The transmission efficiency is calculated by the overlap of the filtered and unfiltered JSA. In other words, only 'intrinsic' losses are considered and experimental limitations such as additional absorption by optical components or absorption losses in the crystal are not taken into account. This is permissible since such experimental losses are constant over the wavelength range. 

The introduction of wavelength-independent losses does not chance the position of the optimum, as it only reduces the transmission efficiency. Wavelength-dependent losses can be understood as an additional filter.

Realistic SPDC settings are guaranteed by constraining the crystal length and pump bandwidth values in the optimizer. The crystal lengths are bounded by what is currently commercially available. The pump bandwidth is bounded to a maximum of roughly $25\,$fs ($\Delta f \approx 17\,$THz) pulses. Such pulses can be realized with commercial Ti:Sapph oscillators. See the supplementary materials for the exact bounds and further details. Furthermore we consider Gaussian-shaped and rectangular-shaped bandpass filters. Rectangular filters are a reasonable approximation of broadband bandpass filters.

In the calculations, only the herald photon is filtered. Also filtering the other photon reduces the heralding efficiency. Typically the increase in purity is not worth the additional losses, especially if finite transmission efficiency of filters is included.

\section{\label{Results}Results}
We now proceed by using the metric of \cite{renema_2018_ArXiv} to compute the optimal filter bandwidth, pump bandwidth and crystal length for KTP, BBO and KDP sources. The upper bound for the error of the classical approximation (Eq. \ref{eq:eqBound}) is set on the conventional $E=0.1$. 

Figure \ref{fig:fig3_results}a) is a parametric plot of the source quality $\alpha$. The transmission efficiency $\eta$ is shown on the y-axis and signal-signal photon indistinguishability $x^2$ on the x-axis. The ideal boson-sampling experiment is located at the top right. Each point represents an optimal SPDC configuration that maximizes $\alpha$ for that crystal corresponding to a fixed filter bandwidth. The black dashed isolines indicate the maximum number of photons $k$ one can interfere, i.e., they are solutions of Eq. \ref{eq:eqBound} for a fixed $E$ and $\alpha$. The weak-filtering regime is in the top left, and the strong-filtering regime is in the bottom right.

Figure \ref{fig:fig3_results}b) represents the source quality $\alpha$ from Fig. \ref{fig:fig3_results}a) explicitly as a function of the filter bandwidth. The left y-axis indicates the source quality $\alpha$. The right y-axis shows the corresponding maximal number of photons $k$. Both graphs show that there is a filter bandwidth that maximizes $\alpha$. From this maximal $\alpha_{\rm{opt}}$ the minimal transmission budget $\eta_{\rm TB}$ can be defined
\begin{equation}
\eta_{\rm{TB}} ~\alpha_{\rm{opt}} = \alpha_{\rm{50}},
\end{equation}
where $\alpha_{\rm{50}}$ denotes the required value of $\alpha$ to perform a 50-photon boson-sampling experiment. The transmission budget defines the minimum required transmission efficiency for all other components together. This includes, for instance, non-unity detector efficiencies. The maximal $\alpha_{\rm{opt}}$ for each crystal and the corresponding SPDC settings are shown in table \ref{tab:tabErrorBudget}.

The physical intuition behind the curves in Fig. \ref{fig:fig3_results}a) is the following. In case of weak to no filtering (top left in Fig. \ref{fig:fig3_results}a)), the transmission efficiency is the highest and the spectral purity the lowest. In this weak filtering regime the crystal length and pump bandwidth are such that the JSA is as factorizable as it can be without filtering. This can also be seen in Fig \ref{fig:fig_GVD_purity}. Examples of such JSAs can be found in the appendix. 

If we now increase filtering, we arrive at the regime of moderate filtering, at the center of Fig. \ref{fig:fig3_results}a). While increasing the filtering, the optimal crystal length increases and the optimal pump bandwidth decreases. This results in a relative increase of the transmission efficiency, since the unfiltered JSA is now smaller and 'fits easier' in the filter bandwidth. The filter also smoothens out the JSA side lobes into a two-dimensional approximate Gaussian. This is the regime with the optimal value for $\alpha$. 

In the case of stronger filtering, the losses start to dominate. The optimal strategy in this regime is to make the JSA as small as possible, such that as much of the photons can get through. By doing so, the 'intrinsic' purity, i.e., before filtering, reduces since this configuration does no longer result in a factorizable state. However, this reduction of purity is compensated by the spectral filter. This is shown in Fig. \ref{fig:fig_GVD_purity}, where the 'intrinsic' purity decreases, but the filtered purity increases.

\begin{figure}
\includegraphics{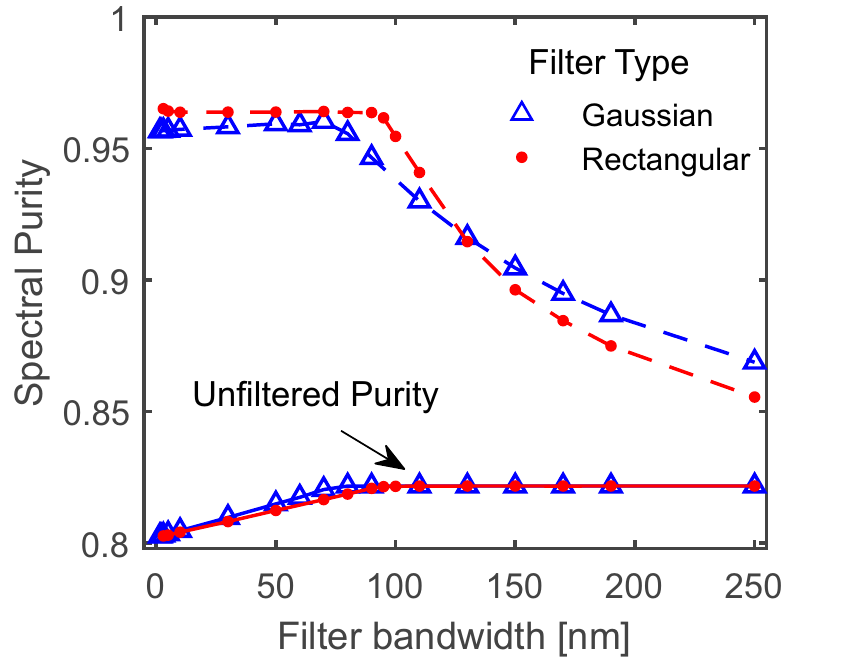}%
\caption{\label{fig:fig_GVD_purity}The spectral purities of a ppKTP source with a sinc phase-matching function. The solid lines describe the spectral purity of the resulting photons before filtering. The dashed lines correspond to the purity after passing through the spectral filter.}
\end{figure}

Furthermore this physical picture also explains the differences between a rectangular and Gaussian filter window. The first difference is that a Gaussian filter allows for higher values of $\alpha$ and thus for more photons in a boson-sampling experiment. The second difference is the optimal filter bandwidth. Both differences can be explained by noting that a rectangular filter window ideally only filters out the side lobes. As a result it cannot increase the factorability of the 'main' JSA, i.e., the part without the side lobes.

The results of the Gaussian apodized source cannot be understood using the aforementioned physical intuition. The filter does not improve the spectral purity since there are no side lobes and the pump bandwidth and crystal length can be chosen such that the JSA is factorizable. The limiting factor here is group-velocity dispersion, which is small around $1582\,$nm \cite{weston_2016_Opt.ExpressOE}.

\begin{table}
\caption{\label{tab:tabErrorBudget}The values of $\alpha_{\rm{opt}}$ and the loss budget for a $k=50$ photon boson-sampling experiment for different crystals at a center wavelength $\lambda_{\rm c}$. The corresponding SPDC settings (crystal length $L$, pump bandwidth $\sigma_{\rm p}$ and filter bandwidth $\sigma_{\rm f}$) are also listed. The mentioned bandwidths are FWHM of the fields.}
\begin{ruledtabular}
\begin{tabular}{l | c c c c c c}
Crystal &$\alpha_{\rm{opt}}$ &$\eta_{\rm{TB}}$ &$\lambda_{\rm c}$ &$L$ &$\sigma_{\rm p}$ &$\sigma_{\rm f}$\\ 
& & &(nm) &(mm) &(nm) &(nm)\\ \hline
KDP &0.9804 &0.8923 &830 &25 &2.3 &6\\
KDP R.\footnote{\label{noteR}Rectangular filter window} &0.976 &0.8964 &830 &25 &2.4 &10\\
ppKTP &0.9051 &0.9667 &1582 &0.5 &21.34 &80\\
ppKTP R. &0.8821 &0.9918 &1582 &0.5 &20.97 &95\\
apKTP &0.9999 &0.8749 &1582 &30 &0.40 &$>$10\\
BBO &0.9106 &0.9608 &1514 &0.95 &30 &110\\
BBO R. &0.8874 &0.9859 &1514 &0.94 &30 &130 \\
\end{tabular}
\end{ruledtabular}
\end{table}

\section{\label{Discussion}Discussion}

It is well known that the spectral purity of symmetrically group-velocity-matched SPDC sources is invariant to changes of either the crystal length or pump bandwidth, as long as the other one is changed accordingly. However, Fig. \ref{fig:fig_GVD_purity} shows that relation no longer holds when filtering is included. In the regime of strong filtering, $\alpha$ is dominated by the losses. Therefore, the SPDC configuration which optimizes $\alpha$ inevitably is the one that minimizes the losses. Hence the spectral purity reduces, but this is compensated by the strong filtering.

In an experiment the non-unity transmission efficiency of a filter at the maximum of the transmission window will be an important source of losses. As a consequence, spectral filtering is only useful when the filter's maximum transmission is larger than $\alpha_{\rm{f}}/\alpha_{\rm{0}}$, where $\alpha_{\rm{f}}$ denotes the filtered $\alpha$ and $\alpha_{\rm{0}}$ the unfiltered case. If the filter's transmission is lower, then the gain in $\alpha$ is not worth the additional losses.

We note that the ideal filter bandwidths of Tab. \ref{tab:tabErrorBudget} are larger than what is reported in \cite{meyer-scott_2017_Phys.Rev.A}. We attribute this difference to two points. Firstly, the model of \cite{meyer-scott_2017_Phys.Rev.A} approximates the sinc phase-matching function as a Gaussian. This eliminates the side lobes and hence reduces the losses. As a consequence, smaller filter bandwidths are optimal. Secondly, the model of \cite{meyer-scott_2017_Phys.Rev.A} focuses on the symmetrized heralding efficiency where both photons are filtered. 

A final point regarding the spectral filters is that the optimal filter bandwidths for ppKTP sources are rather large ($>100\,$nm). Photons with such large bandwidths are typically unpractical for multi-photon experiments since the properties of optical components, e.g., the splitting ratio of a beam splitter, are rarely constant over such a wavelength range. These additional constraints on optical components may result in a better classical simulation of boson sampling. Hence it could increase the required effort to do a boson-sampling experiment.

\section{\label{Conclusion}Conclusion}
In conclusion, we have numerically optimized SPDC sources for scattershot boson sampling. Using the recently found source quality parameter $\alpha$ \cite{renema_2018_ArXiv} we have investigated a number of candidates for building the next generation of SPDC sources. 

From the results of Tab. \ref{tab:tabErrorBudget} we conclude that SPDC sources in principle allow the demonstration of a quantum advantage with boson sampling. The most suitable source for boson sampling is an apKTP crystal. Such a source can have a maximal source quality $\alpha_{\rm{opt}}=0.99$ and has a corresponding transmission budget of $0.87$\%. This transmission budget is sufficient to incorporate state-of-the-art\cite{marsili_2013_NatPhotonics,PhotonSpot95.5} detector efficiencies and keep a small buffer for additional optical losses. The other, periodically poled, KTP source has an optimal filter bandwidth of more than $100\,$nm.

The other two sources are asymmetrically group-velocity-matched sources. The KDP source with a maximal source quality of $\alpha_{\rm opt} = 0.98$ is a good alternative. The optimal source quality for BBO is found to be comparable with ppKTP and less suited for a boson sampling experiment. The fact that these asymmetrically group-velocity-matched sources perform less than symmetrically matches sources is consistent with previous findings.

The limited tolerance for additional losses for the Gaussian apodized KTP source suggests that both waveguide sources and bulk sources without focusing of the pump beam are ideal. Such sources have a single spatial mode and thus do not suffer from an additional reduction of distinguishability which is inevitable with focusing \cite{bennink_2010_Phys.Rev.A}.

This work can be extended to other SPDC sources such as \cite{montaut_2017_Phys.Rev.Applied,jin_2019_Phys.Rev.Applied,laudenbach_2017_Phys.Rev.Applied}, four-wave mixing sources \cite{wang_2001_J.Opt.B:QuantumSemiclass.Opt.} and to Gaussian boson sampling \cite{hamilton_2017_Phys.Rev.Lett.}. The latter can be realized by including the distinguishability between the signal and idler photons.

\begin{acknowledgments}
The Complex Photonic Systems group acknowledges funding from the Nederlandse Wetenschaps Organsiatie (NWO) via QuantERA QUOMPLEX (no. 731473), Veni (Photonic Quantum Simulation) and NWA (No. 40017607). The Integrated Quantum Optics group acknowledges funding from the European Research Council (ERC) under the European Union’s Horizon 2020 research and innovation programme (Grant agreement No. 725366, QuPoPCoRN). 
\end{acknowledgments}

\appendix*
\section{Optimal SPDC settings}
The effect of the filter bandwidth on the optimal SPDC configuration (except for apKTP) can be categorized in three different regimes. These regimes are the weak, moderate and strong filtering regime. An example of the JSA of a ppKTP source in all three regimes can be seen in Fig. \ref{fig:figJSAFilter}. 

\begin{figure*}
\includegraphics{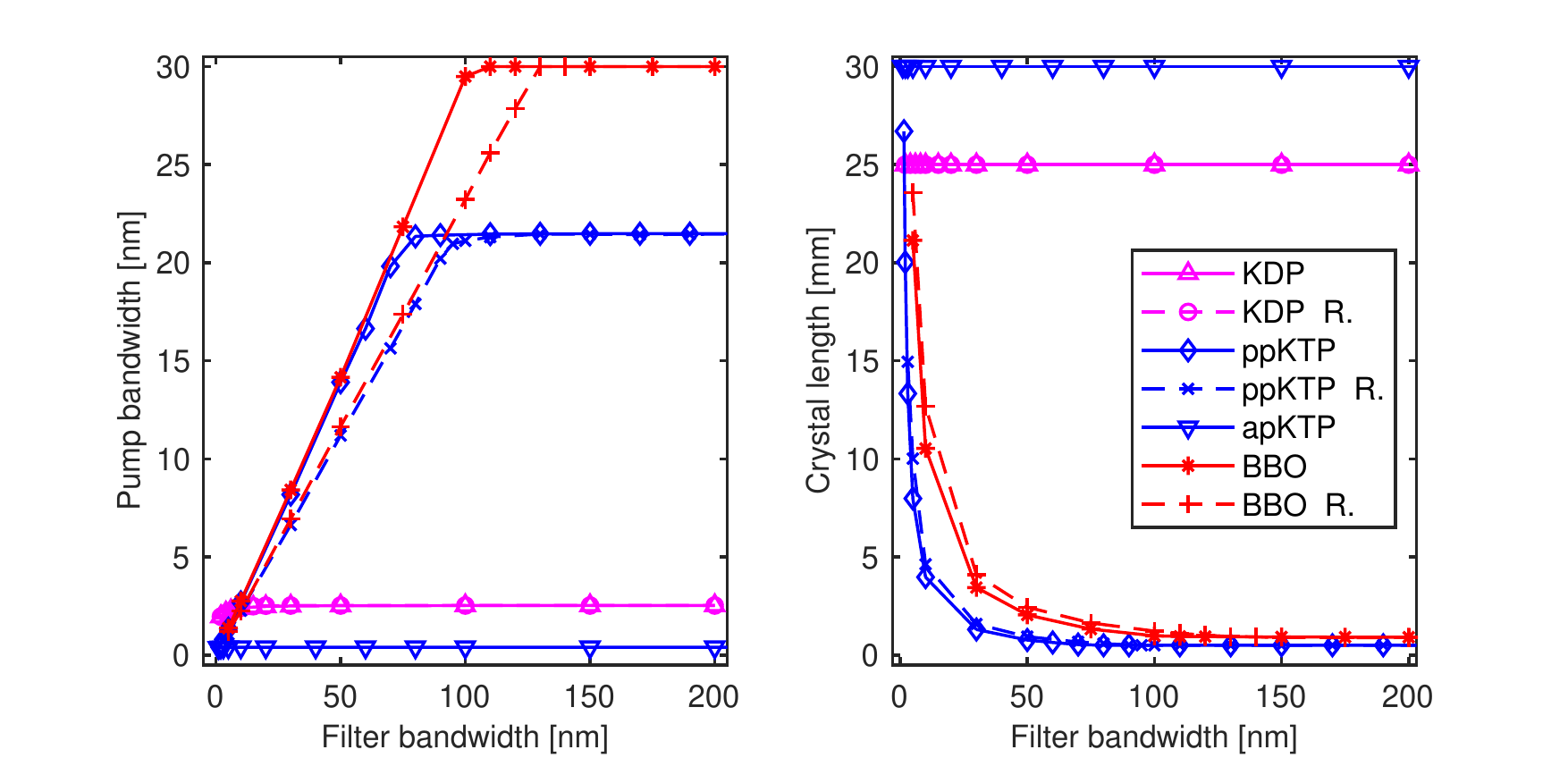}%
\caption{\label{fig:figSettings}The optimal pump bandwidth and crystal length as a function of the filter bandwidth.}
\end{figure*}

\begin{figure*}
\includegraphics{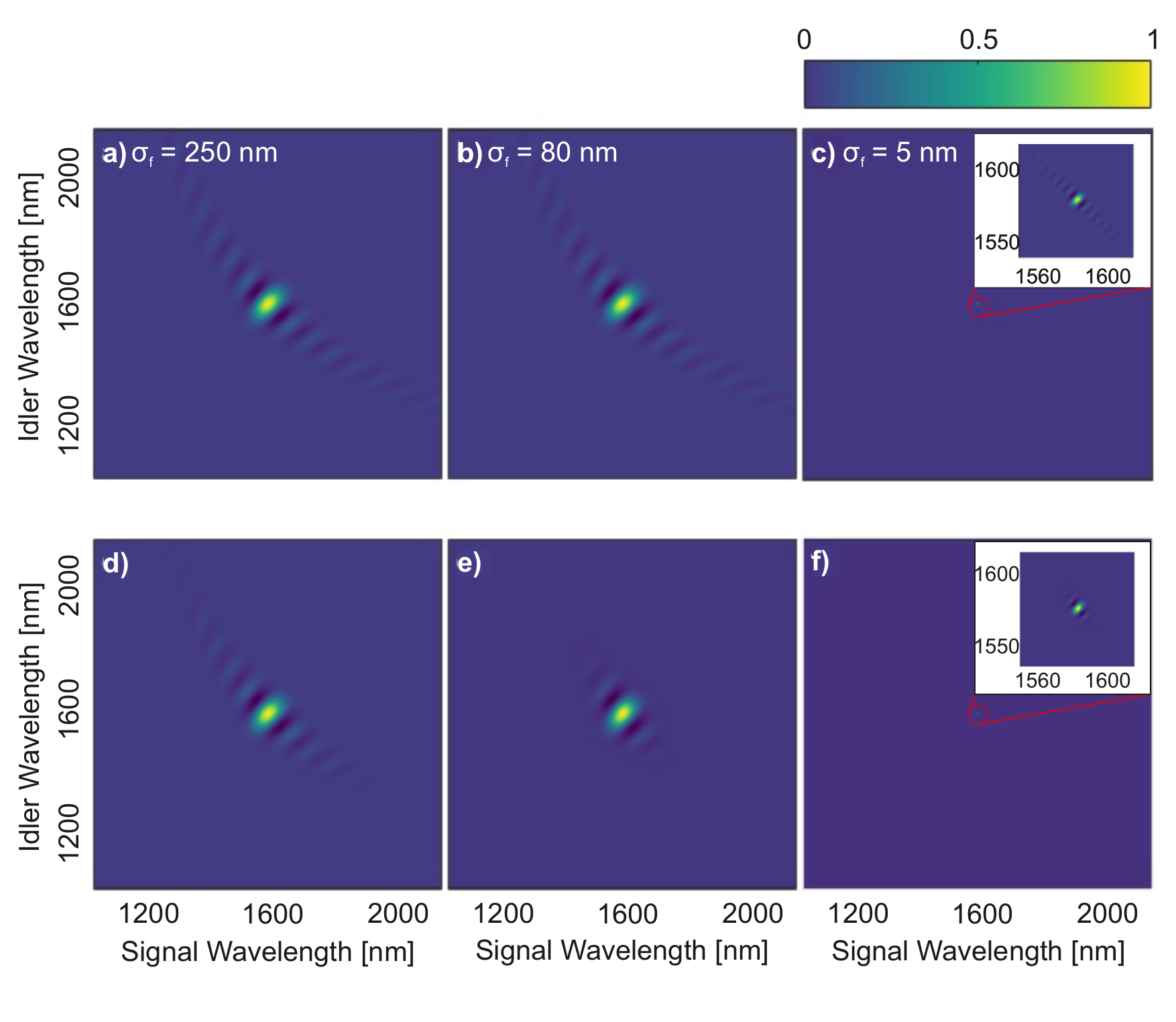}%
\caption{\label{fig:figJSAFilter}The real part of the JSA of the optimal ppKTP SPDC configuration in case of weak filtering with (left), optimal filtering (center) and strong filtering (right). The top panels show the JSA before filtering, the bottom panels after filtering.}
\end{figure*}

The corresponding SPDC configuration parameters can be seen in Figure \ref{fig:figSettings}. This figure shows that in the weak filtering regime, the bounds on the crystal sizes and pump bandwidth can be reached. Once such a bound is reached, the SPDC configuration loses a parameter to optimize the JSA factorizability with, meaning that the general trend of matching the crystal length and pump bandwidth cannot continue anymore. This limits the purity. In case of ppKTP, the limiting factor is the crystal length, whereas in case of a BBO source the maximum pump bandwidth is the limiting factor.


\section{Numerical stability}
We used a local optimization algorithm to find the optimal SPDC configuration for different filter bandwidths. Each iteration of this algorithm computes the spectral purity and losses by discretizing the (filtered) JSA. Such a numerical approach can fail and/or give wrong results. The algorithm can fail because the problem is not convex or that it finds unphysical results (such as a negative crystal length). The algorithm can give wrong results if the discretization of the JSA is too coarse. 

By bounding the parameter space we guarantee that the algorithm does not reach unphysical results. Furthermore, we note that optimizing over the whole parameter space, i.e., the filter bandwidths, crystal lengths and pump bandwidths is not a convex problem. This problem is solved by optimizing the crystal and pump properties each time for different filter bandwidths.

The discretization of the JSA can cause numerical errors. Increasing the number of grid points, i.e., increasing the resolution, decreases this numerical error. Increasing the resolution results to a convergence of the result. Unfortunately, it is not directly known how our numerical calculation converges to a reliable answer. How to a priori estimate the numerical error for a given discretization is also unclear.

\begin{figure}
\includegraphics{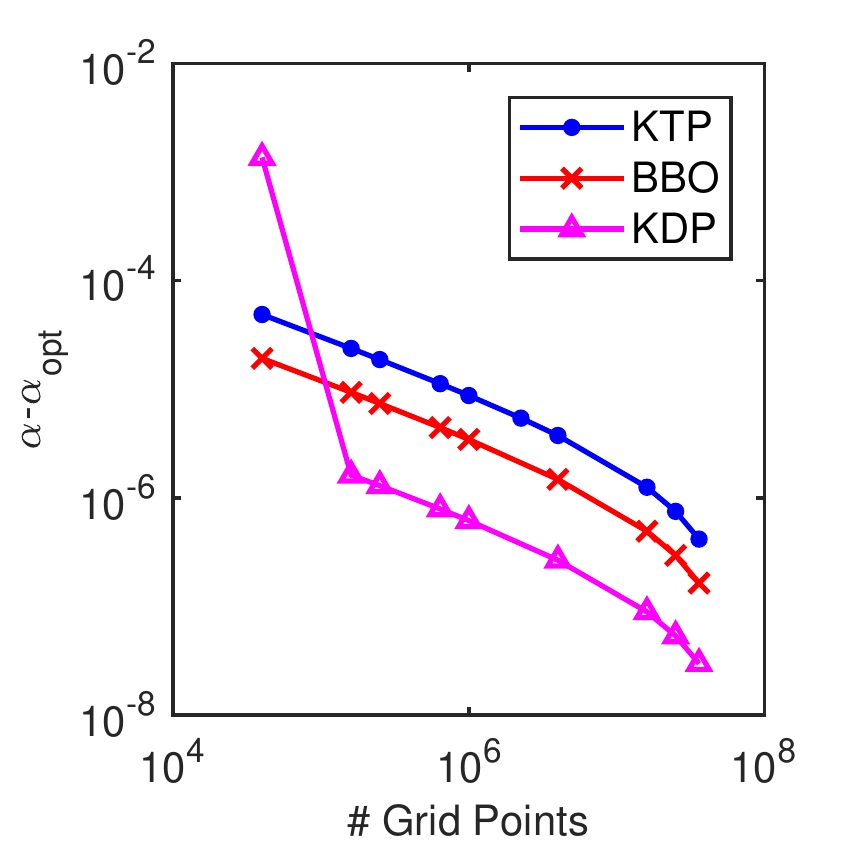}%
\caption{\label{fig:figGridpoints}The convergence of the source quality $\alpha$ with the discretization of the frequency space. Each data point is the difference of $\alpha$ with the $\alpha$ corresponding to $8000^2$ grid points. All crystals, ppKTP, BBO and KDP, are set in their optimal configuration.}
\end{figure}

In order to show that our calculations have converged, we simply try different discretizations of the JSA. For every discretization, we calculate the corresponding source quality $\alpha$ and observe how it is varies. Figure \ref{fig:figGridpoints} shows that the numerical error originating from this discretization is small in the limit of more than $2000^2$ ($2000$ per photon) grid points. This confirms the validity of our calculations. Table \ref{tab:tabSimBounds} provides an overview of all relevant parameters for the stability of the simulation.

\begin{table*}
\caption{\label{tab:tabSimBounds}The simulation parameters for each crystal. The bounds on the crystal lengths and pump bandwidth are given, just as the range of wavelength over which the JSA is computed. The grid points are the number of steps used to discretize the entire wavelength range}
\begin{ruledtabular}
\begin{tabular}{l | c c c c c c c c}
Crystal &\multicolumn{2}{c}{Crystal Length} &\multicolumn{2}{c}{Pump bandwidth} &\multicolumn{2}{c}{Wavelength} &Grid points &Sellmeier constants\\\cline{2-3}\cline{4-5}\cline{6-7}
&minimum &maximum &minimum &maximum &minimum &maximum & &\\
&(mm) &(mm) &(nm) &(nm) &(nm) &(nm) & & \\ \hline
KTP &0.5 &30 &0.1 &30 &1028 &2136 &$2000^2$ &\cite{Konig_2004_Appl.Phys.Lett.,fradkin_1999_Appl.Phys.Lett.}\\
BBO &0.5 &40 &0.1 &30 &1008 &2093 &$2000^2$ &\cite{tamosauskas_2018_Opt.Mater.ExpressOME}\\
KDP &0.5 &25 &0.1 &10 &780 &880 &$1500^2$ &\cite{menzel_2013_} \\
\end{tabular}
\end{ruledtabular}
\end{table*}


\bibliography{SPDC_paper_ref}

\end{document}